\documentclass{article}

\usepackage[preprint,nonatbib]{neurips_2022}

\usepackage[utf8]{inputenc} 
\usepackage[T1]{fontenc}    
\usepackage{hyperref}       
\usepackage{url}            
\usepackage{booktabs}       
\usepackage{multirow}
\usepackage{amsfonts}       
\usepackage{nicefrac}       
\usepackage{microtype}      
\usepackage{xcolor}         
\usepackage{graphicx}
\usepackage{floatrow}
\newfloatcommand{capbtabbox}{table}[][\FBwidth]

\title{Modeling Animal Vocalizations through Synthesizers}

\author{%
  Masato Hagiwara \\
  Earth Species Project \\
  \texttt{masato@earthspecies.org} \\
   \And
    Maddie Cusimano \\
    Earth Species Project \\
  \texttt{maddie@earthspecies.org}
   \And
    Jen-Yu Liu \\
  Earth Species Project \\
  \texttt{jenyu@earthspecies.org} \\
}

\begin{document}

\maketitle




\begin{figure}[h]
\begin{center}
\includegraphics[scale=0.38]{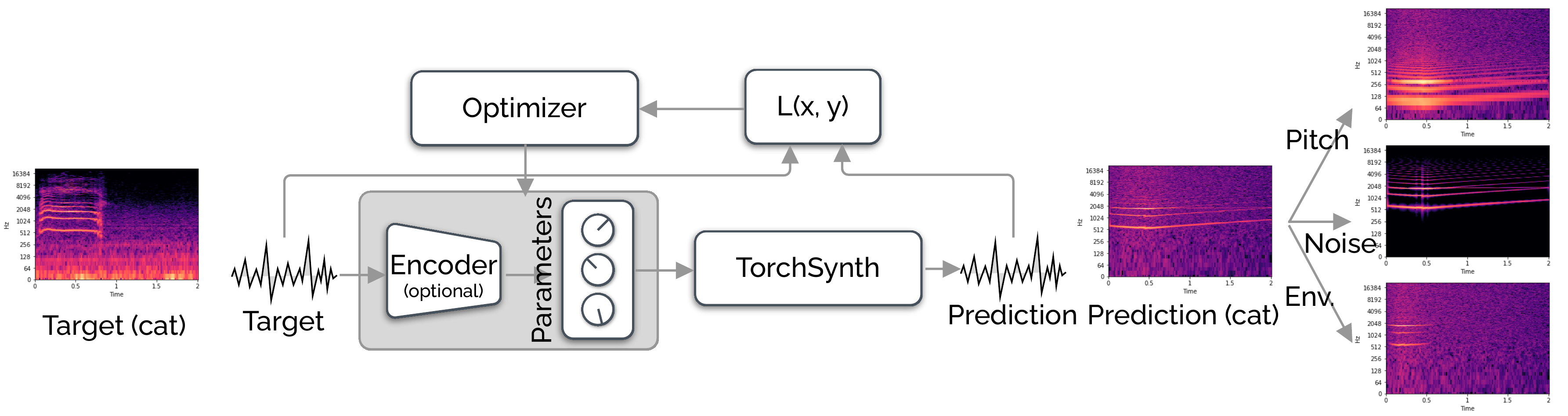} 
\caption{Method overview. Given an animal sound, synthesizer parameters are optimized for reconstruction. The predicted sound can be modified (e.g. pitch-shift, denoising, envelope modulation).}
\label{fig:overview}
\end{center}
\end{figure}

\section{Introduction}

Modeling real-world sound is a fundamental problem in the creative use of machine learning and many other fields, including human speech processing and bioacoustics. Transformer-based generative models~\cite{li2019neural,dhariwal2020jukebox} are known to produce realistic sound, although they have limited control and are hard to interpret. Recently, lighter-weight models that incorporate structured modules and domain knowledge, notably DDSP~\cite{engel2020ddsp,wu2022mididdsp}, have been shown to produce high-quality musical sound. However, a lack of signal-processing knowledge may hinder users from effectively manipulating the synthesis parameters, of which there can be over a hundred per frame.

As an alternative, we aim to use modular synthesizers, i.e., compositional, parametric electronic musical instruments, for modeling non-music sounds\footnote{Audio samples are at \url{https://earthspecies.github.io/animalsynth/}}. Synthesizers are lightweight and designed in part for control, which make them a plausible candidate model. However, inferring synthesizer parameters given a target sound, i.e., the {\it parameter inference} task~\cite{justice1979analytic,garcia2002automatic}, is not trivial for general sounds. Research utilizing modern optimization techniques has typically focused on musical sound~\cite{yeeking2018automatic,esling2020flow,le2021improving,barkan2019inversynth,chen2022sound2synth}. In this work, we optimize a differentiable synthesizer from TorchSynth~\cite{turian2021torchsynth} in order to model, emulate, and creatively generate animal vocalizations. We compare an array of optimization methods, from gradient-based search to genetic algorithms, for inferring its parameters, and then demonstrate how one can control and interpret the parameters for modeling non-music sounds. The current work is intended as a creative tool for artists to integrate animal sound into their work, but we hope future work in this direction could also enable new synthesis methods in bioacoustics~\cite{king2015you}.

\section{Parameter Inference}

\begin{figure}
\begin{floatrow}

\capbtabbox{%
{\small
\begin{tabular}{lrr}
\toprule
Method           & Loss ($\downarrow$)  & Acc.  ($\uparrow$) \\ \midrule
Random search    & 0.861                    & 0.068       \\
Gradient (Adam)  & 1.553                    & 0.027       \\ 
Variational optimization  & 0.891                    & 0.034       \\ 
Genetic algorithm & 0.633                   & 0.114       \\
Differential evolution & 0.738               & 0.068       \\
PGPE \cite{sehnke2010parameter}             & 1.000                    & 0.011       \\
CMA-ES \cite{hansen2001evolutionary}           & 0.891                    & 0.027       \\
Metropolis MCMC  & 0.658                    & 0.091       \\ 
Bayesian optimization \cite{ozaki2020multiobjective} &   0.737         &  0.034         \\
Encoder          & 2.131                    & 0.034       \\
Original         & ---                      & 0.841       \\
\bottomrule
\end{tabular}
}
}{%
    \caption{Comparison of optimization methods}
    \label{table:optimization}
}

\ffigbox{%
    \includegraphics[scale=0.38]{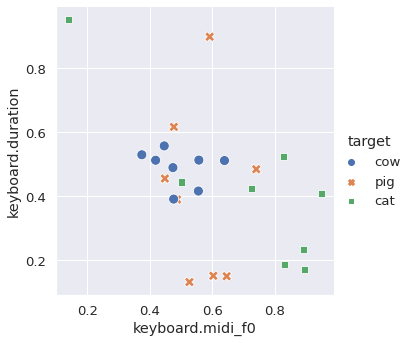} 
}{%
    \caption{Plot of parameters per category. Parameters are normalized to the [0,1] range.}
    \label{fig:plot}
}

\end{floatrow}
\end{figure}

The default synthesizer implemented in TorchSynth named {\it Voice} has a relatively simple architecture with a total of 78 parameters, consisting of two voltage-controlled oscillators (VCOs) and a noise generator, whose pitch and amplitude envelopes are modulated by a mix of low-frequency oscillators (LFOs). For any non-trivial synthesizer, it is extremely difficult to adjust parameters manually to reproduce the target sound, necessitating automated optimization techniques.

We first compare several optimization techniques for parameter inference. As the target sound dataset, we used the animal sounds (e.g., dog, cat, birds) in the first fold of ESC-50~\cite{piczak2015dataset}, an environmental sound dataset. We used the multi-scale STFT loss implemented in auraloss~\cite{steinmetz2020auraloss} as the objective. In order to assess the quality of reconstruction, we fine-tuned the pretrained VGGish~\cite{hershey2017cnn} model on the remaining four folds of ESC-50 on the 50-class sound classification task (which includes both animal and non-animal categories) and used its accuracy. Intuitively, if an optimization method achieves good reconstruction, it can also ``fool'' the 
classifier trained on real sounds.

Table~\ref{table:optimization} shows the optimization techniques and their metrics. See the Supplemental Information for the details of individual methods. Overall, evolutionary algorithms (the genetic algorithm and differential evolution) worked relatively well, while gradient-based methods performed poorly, due to the fact that the objective function is highly nonlinear and complex with respect to synthesizer parameters. This trend is also observed for other highly complex optimization problems~\cite{tian2022modern}. Predicting parameters directly with a neural network encoder was not successful either, possibly due to the discrepancies between the real sounds and the training data (artificially generated sound by TorchSynth itself). 

One potential way to improve the gradient-based methods is to use a proxy network that simulates the synthesizer, and replace the original synthesizer with the proxy network in parameter inference~\cite{steinmetz2020effect}. We trained a MelGAN~\cite{kumar2019melgan}-based proxy network that takes the TorchSynth parameters as input and the corresponding waveform as the target. We found that the reconstructions were poor. In contrast to audio effects, converting 78 parameters to a 2-second waveform (88,200 samples) may have its own challenges due to the large difference in the scales of the input and output.

After inferring parameters for each sample/class, controlling and generating new sounds are trivial. For example, one can change pitch, de-noise, and/or modify the amplitude envelope by changing just one or few synthesizer parameters as shown in the spectrograms in Figure~\ref{fig:overview}.

\section{Interpretation and Generation}

One benefit of synthesizer parameter inference is that the fitted parameters are interpretable. In Figure~\ref{fig:plot}, we choose three animal categories (cow, pig, cat) for which the classifier accuracy is relatively high and plot the fundamental frequency (f0) and the duration of the note inferred by the genetic algorithm. The sound instances cluster based on their acoustic properties, i.e., cats make short and high-pitched sounds while cow vocalizations are longer and lower-pitched.

Finally, as a proof of concept, we generated new cat sounds by first fitting a Gaussian distribution over the inferred cat parameters (green squares in Figure~\ref{fig:plot}) in the unnormalized space, then generating 100 new sounds by sampling from this distribution. The fine-tuned VGGish model correctly classified 18 samples out of 100, demonstrating that we can generate plausible new sounds by sampling synthesizer parameters from the fitted distribution.

\section*{Ethical Considerations}
With the progress of audio synthesis technology in recent years, synthesized human voices could be used in harmful activities \cite{meskys2020deepfake,diakopoulos2021deepfake,lee2022ethics}. Similarly, playing synthesized animal sounds in the wild could also be harmful to animals \cite{cuthill1991field,harris2013playback}, meaning released models could have potential negative effects. One way to mitigate such risks is to accompany the generative models with corresponding detection models \cite{borsos2022audiolm}.

With the above considerations, simpler models (like the one used in this work) could be a good alternative to more powerful generative models to be released for creative use. Its constrained synthesis makes it difficult to synthesize authentic sounds, while still capturing evocative aspects of the target sounds and providing rich variations to be used in creative design.

\section*{Acknowledgement}

The authors would like to thank Benjamin Hoffman, Felix Effenberger, and Katie Zacarian at Earth Species Project for their valuable feedback on this project.

\bibliography{neurips_2022}
\bibliographystyle{plain}

\section*{Supplemental Information}

\subsection*{Parameter Inference}

\textbf{Model.} We used the default synthesizer implemented in TorchSynth, named \textit{Voice}, with the default `voice' nebula. See \url{https://torchsynth.readthedocs.io/en/latest/modular-design/modular-principles.html##synth-architectures} for details.

\textbf{Target dataset.} ESC-50 \cite{piczak2015dataset} contains ten animal sound classes: dog, rooster, pig, cow, frog, cat, hen, insects flying (e.g., buzzing), sheep, crow, and birds. To create the target sound dataset, we took the subset of sounds in the first fold that were in these ten classes. All sounds were two seconds and sampled at 44.1 kHz. 

\textbf{Reconstruction loss.} We used a mel-scaled multi-scale short-time Fourier transform (STFT) loss from the auraloss package \cite{steinmetz2020auraloss}, with equally weighted linear-scale magnitude and spectral convergence loss terms. Four resolutions had FFT, hop and window sizes corresponding to (2048, 1024, 512, 256), (256, 128, 64, 32) and (1024, 512, 256, 128) respectively. All resolutions used 45 mel-frequency bins. 

\subsection*{Optimization}

Synthesizer parameter were initialized with uniform random samples, unless otherwise noted. Hyperparameters were manually chosen. 

\textbf{Random search.} We took 1000 random uniform samples from all 78 parameters. For each sound, we selected the sample that minimized the reconstruction loss.

\textbf{Gradient (Adam).} We used 200 iterations of an adaptive gradient descent algorithm (Adam) on all 78 parameters, with a learning rate of 0.001. For each sound, we selected the parameters on the final iteration. 

\textbf{Variational optimization.} The original 78 parameters of the modular synthesizer were used in conjunction with an additional 78 variance parameters to define a Beta distribution over each synthesizer parameter. The distributions were initialized to uniform and then optimized with Adam for 500 iterations and a learning rate of 0.001, using a batch size of 200 samples. For each sound, we selected the sample from the last iteration that minimized the reconstruction loss.

\textbf{Genetic algorithm.} We used the pygad package \cite{gad2021pygad}. For 100 iterations of steady-state selection, four parents were selected and single-point crossovers and random mutations (10\% of parameters) were used to evolve the parameters. For each sound, we selected the individual that minimized the reconstruction loss.

\textbf{Differential evolution.} We used the default differential evolution function from the scipy package, with a maximum of 20 generations and a population size of 10. For each sound, we selected the individual that minimized the reconstruction loss.

\textbf{Policy gradients with parameter-based exploration (PGPE).} We use PGPE implemented in the pgpelib package \cite{toklu2020clipup}, a distribution-based evolutionary algorithm that estimates gradients with samples around a center solution. We use a population size of 100 evolved over 100 generations using the ClipUp optimizer. For each sound, we use the center solution of the final generation.  

\textbf{Covariance matrix adaptation Evolutionary Strategy (CMA-ES).} We used CMA-ES as implemented in the pycma package \cite{hansen2001evolutionary}, with a maximum of 200 iterations. For each sound, we use the mean solution. 

\textbf{Metropolis algorithm.} We implemented Markov Chain Monte Carlo (MCMC) using the Metropolis algorithm. The synthesizer was supplemented with a uniform prior for each parameter and a Gaussian noise model ($\sigma = 0.1$) for the likelihood. On each iteration, we randomly selected parameters to be resampled ($p=0.1$) with a uniform random proposal function. We collected 10000 posterior samples. For each sound, we selected the posterior sample that minimized the reconstruction loss. 

\textbf{Bayesian optimization.} We used the default hyperparameter optimization in the optuna package \cite{takuya2019optuna} (Tree-Structured Parzen Estimator \cite{ozaki2020multiobjective}), with 1000 trials.

\textbf{Encoder.} We fine-tuned the pretrained VGGish~\cite{hershey2017cnn} with a fully-connected regression layer on top to predict the synthesizer parameters from the spectrogram. The model is trained with an MSE loss on a dataset of 60,000 sounds generated from TorchSynth.

\subsection*{Generation}

The Gaussian distribution over the inferred cat parameters was fit in the unnormalized space (keyboard.duration in sec, keyboard.midi\_f0 in midi-values). 

\end{document}